\documentstyle[12pt]{article}
\def\dspace{\baselineskip=0.3 in}
\begin{document}
\dspace

\centerline{\bf Non-violation of Energy Conditions in the future}
\centerline{\bf  accelerated universe due to quantum effects }

\vspace{1cm}

\centerline{\bf S.K.Srivastava,}
\centerline{\bf Department of Mathematics,}

\centerline{\bf North Eastern Hill University,}

\centerline{\bf Shillong - 793022}

\centerline{\bf ( INDIA )}

\centerline{\bf e - mail : srivastava@nehu.ac.in; sushil@iucaa.ernet.in}

\vspace{2cm}

\centerline{\bf Summary}

Here, an accelerated phantom model for the late universe is explored, which is
free from future singularity. It is interesting to see that this model
exhibits strong curvature for all time in future, unlike models with
`big-rip singularity' showing high curvature near singularity time
only. So, quantum gravity effects grow dominant as time increases in
late universe too. More importantly, it is demonstrated that quantum
corrections to FRW equations lead to non-violation of `cosmic energy
conditions' of general relativity, which are violated for accelerating
universe without these corrections.

\newpage

Observations [1-3], around turn of the last century, indicate that
we are living in a spatially flat and accelerated universe. Cosmic
acceleration is endowed with $dark$ $energy$ having $-1 \le  {\rm w} = p/\rho
< - 1/3$ with energy density $\rho > 0 $ and pressure $p < 0$. With the
equation of state parameter $-1 \le {\rm w} < - 1/3$, `` strong energy condition ''(SEC) is
violated. Later on, Caldwell advocated for ${\rm w} < - 1$, which violates
`` weak energy condition '' (WEC) also and proposed $phantom$ $model$. This
model ends with singularity in finite future time [4] with
big-smash. A comprehensive review on dark energy is available
in[5]. In [6],Barrow  emphasized that, to get finite future-time singularity,
violation of WEC is not necessary for accelerated universe. This model is
generalized in [7]. Lake [8] has shown
that such cosmologies violate `dominant energy condition' (DEC), but satisfy
`null energy condition' (NEC), SEC and WEC.

In phantom models, having ``big-rip singularity'',  curvature
 becomes very strong near $t_s$.  In curved space-time, quantum corrections contain
 higher order terms of curvature. So, quantum gravity effects become
 dominant for a small interval $|t_s - t| <$ one 
unit of time ( with $t$ being the cosmic time measured in units GeV$^{-1}$) [7,9,10,11]. It is similar to  models of the early
 universe, where curvature invariants are very strong. The role of
 quantum gravity for entropy, in phantom universe, is discussed in
 [12]. In [13], avoidance of big-rip problem is shown without using
 quantum corrections. In [14], using modified gravity, it is shown
 that big-rip problem does not arise in the model with curvature dark energy.
 
Here, it is shown that, for curvature invariants to grow sufficiently strong,
future singularity is sufficient but not necessary. In what follows, a phantom
model of the accalerated universe  is explored,
where quantum effects grow dominant. This model is free from four
types of future singularities mentioned in [15]. In this model, it is
shown that quantum corrections in Friedmann equations lead to {\em
non-violation} of `cosmic energy conditions' even in accelerated
universe. Here, natural units ($\hbar = c = 1$) are used with Gev as the fundamental unit.

Having compatibility with experimental evidences [1-3], topology of the observable universe is given by

$$ dS^2 = dt^2 - a^2(t) [ dx^2 + dy^2 + dz^2 ]. \eqno(1)$$
Based on ``cosmological principle'', stress-energy tensor for content of the universe has perfect fluid form. So, energy conditions are given as[8]

$$ {\rm NEC} \Longleftrightarrow \rho + p \ge 0 , \eqno(2)$$ 

$$ {\rm WEC} \Longleftrightarrow \rho \ge 0 , \quad\rho + p \ge 0 , \eqno(3)$$ 

$$ {\rm SEC} \Longleftrightarrow \rho \ge 0 , \quad \rho + 3 p \ge 0 , \eqno(4)$$
and

$$ {\rm DEC} \Longleftrightarrow \rho \ge 0 , \quad \rho \pm  p \ge 0 . \eqno(5)$$

The action for the phantom scalar $\phi$ is taken as
$$ S = \int {d^4x} \sqrt{-g} \Big[ - \frac{1}{2}
g^{\mu\nu}\partial_{\mu}{\phi}\partial_{\nu}{\phi} - V(\phi)
\Big], \eqno(6a)$$
where potential
$$ V(\phi) = V_0 + \frac{1}{2} m^2 \phi^2  \eqno(6b)$$
with $m$ being the mass.

So, in the homogeneous universe given by eq.(1), energy density $\rho$ and pressure $p$ are obtained as

$$ \rho = - \frac{1}{2} {\dot \phi}^2 + V_0 + \frac{1}{2} m^2 \phi^2 \eqno(7)$$
and
$$ p = - \frac{1}{2} {\dot \phi}^2 - V_0 - \frac{1}{2} m^2 \phi^2  \eqno(8)$$

Moreover, action (6a) and eq.(6b) yield field equation for $\phi$
as

$$ - {\ddot \phi} - 3 H {\dot \phi} + m^2 \phi = 0,
 \eqno(9)$$
where $H = {\dot a}/{a}.$

Friedmann equations are obtained as
$$ \frac{3 H^2}{8 \pi G} = \rho = - \frac{1}{2} {\dot \phi}^2 + V_0
+\frac{1}{2} m^2 \phi^2   \eqno(10a)$$ 
using $\rho$ given by eq.(7)
and
$$  \frac{3 }{4 \pi G} \Big(\frac{\ddot a}{a} \Big) = 2({\dot \phi}^2 + V_0
+\frac{1}{2} m^2 \phi^2  ),  \eqno(10b)$$
where $ G = M_P^{-2} ( M_P = 10^{19} {\rm GeV}$ is the Planck
mass). Conservation equation is 

$$ {\dot \rho} + 3 H (\rho +  p) = 0 . \eqno(11)$$

With $\rho$ and $p$ from eqs.(7),(8), eqs.(11) yield eq.(9). Eqs.(10a) and
(11) give eq.(10b), so eqs.(10a) and (10b) are not independent like eqs.(9)
and (11). Hence, it is enough to solve  only eqs. (9) and (10a).

Eqs.(10a) and (9) integrate to a 
well-behaved scale factor
$$ a(t) = a_0 exp[\frac{m \phi_0\sqrt{12 \pi G}}{3}(t - t_0)+ \frac{m^2}{6}(t
-  t_0)^2 ]   \eqno(12a)$$  
and
$$ \phi = \phi_0 + \frac{m}{\sqrt{12 \pi G}}(t - {\bar t}_0) \eqno(12b)$$
with
$$ V_0 = \frac{m^2}{24 \pi G}, \eqno(13a)$$  
$$ m \phi_0 = H_0 \sqrt{\frac{3}{4 \pi G}} \eqno(13b)$$
and $t_0 = 13.7 {\rm Gyr} = 6.6 \times 10^{41} {\rm GeV}^{-1}$ being the
present age of the universe. Here $H_0 = 2.32 \times 10^{-42} h {\rm GeV}$ is
the current Hubble's rate.

Moreover, eq.(12a) yields

$$ \frac{\ddot a}{a} > 0 \eqno(14)$$
showing speeded-up expansion. Eqs. (10a) and (10b) imply
$$ \rho > 0 \eqno(15a)$$
and
$$ \rho + 3 p  < 0 .\eqno(15b)$$
Also, from eqs.(7),(10a,b,c) and (12a,b), it is obtained that

$$ \rho + p = - \frac{\dot H}{4 \pi G} = - \frac{m^2}{12 \pi G} < 0 \eqno(15c)$$
and
$$ \rho - p =  \frac{{\dot H} + 3 H^2}{4 \pi G} > 0 \eqno(15d)$$

Thus, through classical approach, we get violation of all energy conditions
mentioned in eqs.(2)-(5).

Now, effect of quantum corrections to eqs.(10b), (15c) and (15d) is discussed.

Eq.(12b) yields
$$ H (t) = \frac{m \sqrt{12 \pi G}}{3} \Big[ \phi_0 + \frac{m}{\sqrt{12 \pi
    G}}(t - t_0) ] . \eqno(15e)$$
This equation shows that curvature increases with time $t >  t_0$ for expansion
    with no future singularity. Moreover, eqs.(7) and (12a,b) show growth of phantom energy density also.
     So, as discussed above, quantum effects are expected to be dominant for
    $t > t_0  $ due to increase in curvature. It is unlike models having ``big-rip singularity'', where
    quantum effects dominate only for a small time interval near singularity [7,9,10,11]. 

Upto adiabatic order 4, one-loop quantum corection to the action (6) is given
by[16]

\begin{eqnarray*}
\Gamma &=& \int {d^4 x} \sqrt{ - g} \Big[\frac{3 m^4}{4} - \frac{m^2}{6} R +
ln\Big(\mu^2/m^2 \Big) \Big(\frac{m^4}{2} - \frac{m^2}{2} R +
\frac{1}{30}{\Box}R \\&& + \frac{1}{180} R^{\mu\nu\alpha\beta}
R_{\mu\nu\alpha\beta} - \frac{1}{180} R^{\mu\nu}R_{\mu\nu}+ \frac{1}{72}
R^2  \Big]. 
\end{eqnarray*}
$$ \eqno(16)$$

So, quantum corrections to stress tensor is obtained as

\begin{eqnarray*}
T^q_{\mu\nu} &=& - \frac{m^2}{3} \Big[ 1 + ln\Big(\mu^2/m^2 \Big) \Big]
(R_{\mu\nu} - \frac{1}{2}g_{\mu\nu} R) + 2 ln\Big(\mu^2/m^2 \Big) \Big[
\frac{1}{30} ({\Box}R_{\mu\nu} \\&& - \frac{1}{2}g_{\mu\nu} {\Box}R) -
\frac{1}{2} g_{\mu\nu} R^{\alpha\beta\gamma\delta} R_{\alpha\beta\gamma\delta}
+ 2 R_{\mu\alpha\beta\gamma} R^{\alpha\beta\gamma}_{\nu} - 4 {\Box}R_{\mu\nu}
+ 2 R_{;\mu\nu} \\&&  - 4 R_{\mu\alpha}R^{\alpha}_{\nu}  - 4 R^{\alpha\beta}
R_{\alpha\mu\beta\nu})  \\&& - \frac{1}{180}( 2 R^{\alpha}_{\mu; \alpha\nu} - {\Box} R_{\mu\nu} -
\frac{1}{2}g_{\mu\nu} {\Box}R + 2 R^{\alpha}_{\mu} R_{\alpha \nu} \\&& -
\frac{1}{2}g_{\mu\nu} R^{\alpha\beta} R_{\alpha\beta}) + \frac{1}{72}(2
R_{;\mu\nu} - 2g_{\mu\nu} {\Box}R - \frac{1}{2}g_{\mu\nu} R^2 + 2 R R_{\mu\nu}
)\Big]
\end{eqnarray*}
$$ \eqno(17)$$
leading to
\begin{eqnarray*}
\rho^q = T^{0(q)}_0 &=& m^2 \Big[ 1 + ln\Big(\mu^2/m^2 \Big) \Big] H^2 + 2
ln\Big(\mu^2/m^2 \Big) \Big[ - \frac{13}{10} H {\ddot H} \\&& - \frac{23}{40} {\dot
  H}^2 - \frac{13}{5} {\dot H} H^2 - \frac{6}{5} H^4 \Big]
\end{eqnarray*}
$$ \eqno(18a)$$
and
\begin{eqnarray*}
-p^q &=& T^{1(q)}_1 = T^{2(q)}_2 = T^{3(q)}_3 \\ &=& \frac{ m^2}{3} \Big[ 1 +
ln\Big(\mu^2/m^2 \Big) \Big](2 {\dot H} + 3 H^2) + 2
ln\Big(\mu^2/m^2 \Big) \Big[ - \frac{9}{40} \frac{d{\ddot H}}{dt} \\&& +
\frac{29}{120} H {\ddot H} + \frac{31}{40} H {\dot H} + \frac{5}{6} {\dot
  H}^2 + \frac{4}{5} {\dot H} H^2 + \frac{6}{5} H^4 \Big]
\end{eqnarray*}
$$ \eqno(18b)$$

Using $H(t)$,given by eq.(16), eqs.(18a,b) look like
$$\rho^q = m^2 \Big[ 1 - \frac{11}{15} ln\Big(\mu^2/m^2 \Big) - \frac{12}{5
  m^2}  ln\Big(\mu^2/m^2 \Big) H^2 \Big] H^2 - \frac{23}{180} m^4
  ln\Big(\mu^2/m^2 \Big) \eqno(19a)$$
and
\begin{eqnarray*}
-p^q &=&  \frac{ m^2}{3} \Big[ 1 +
ln\Big(\mu^2/m^2 \Big) \Big](2 \frac{m^2}{3} + 3 H^2) + 2
ln\Big(\mu^2/m^2 \Big) \Big[  \frac{5}{54} m^4 \\&& + \frac{31}{120} m^2
  H + \frac{4}{15}m^2 H^2 + \frac{6}{5} H^4 \Big]
\end{eqnarray*}
$$ \eqno(19b)$$

Eqs.(16-19b) show that $H$, given by eq.(15c), dominates quantum
corrections. So, a drastic change, in the behaviour of matter, is obtained
with the same geometry (with $a(t)$ from eq.(12b)) on replacing $\rho$
by $\rho + \rho^q$ and $p$ by $p + p^q$ in eqs.(10b),(15c) and (15d). These
corrections yield  
 
\begin{eqnarray*}
\rho + 3 p &=& - \frac{3}{8 \pi G} \Big(\frac{m^2}{3} + H^2 \Big) - \rho^q - 3
p^q \\&& \approx \Big[ - \frac{3 M^2_P}{8 \pi } + \frac{364 \pi m^2}{45
  M^2_P}ln\Big(\mu^2/m^2 \Big) \Big\{ \phi_0 + \frac{m M_P}{\sqrt{12 \pi
    }}(t - t_0) \Big\}^2 \Big] H^2, \\ \rho + p &=& -
\frac{m^2}{12 \pi G} - \rho^q - p^q \approx  \frac{24}{5}ln\Big(\mu^2/m^2
\Big) H^4 > 0, \\ \rho - p &=& 
\frac{m^2}{12 \pi G} + \frac{3 H^2}{4 \pi G}- \rho^q + p^q \approx
 \Big[  \frac{3 M^2_P}{4 \pi } - m^2 \Big(1 + ln\Big(\mu^2/m^2 \Big) \Big] H^2
\end{eqnarray*}
$$ \eqno(20a,b,c)$$
as terms containing $H^4$ dominate. Here  $G = M_P^{-2}$ is used.

Eq.(20a) shows that $\rho + 3 p > 0$ if
$$  \frac{3 M^2_P}{8 \pi } < \frac{91}{45  }ln\Big(\mu^2/m^2 \Big) \Big\{ 3
H_0 + m^2 (t - t_0) \Big\}^2 \eqno(21)$$
using $\phi_0$ from eq.(13b).

Setting $\mu = m \sqrt{e} ( e$ stands for exponential) in eq.(21), it turns
out that for $t > t_0 + \frac{0.243 M_P}{m^2}, \rho
  >0 , \rho \pm p > 0 $ and $\rho + 3 p> 0$. It shows that energy conditions,
  violated without accounting for quantum effects, are restored using quantum
  corrections when $t > t_0 + \frac{0.243 M_P}{m^2}$.

Thus, it is obtained that a phantom model, without `future singularity' and
exhibiting  accelerated expansion,  brings
back quantum gravity era for a long span of time in future, after early
universe  and subsequent development of the universe upto
the present epoch. It is unlike models having `big-rip singularity', where
these effects are dominant for a short time. Interestingly, here, restoration of energy
conditions SEC, WEC, DEC and NEC is possible on taking quantum effects into
account, which are violated without taking these corrections for a speeded-up
universe. It is due to negative $\rho^q$ and negative $p^q$ as quantum corrections,
given by eqs.(18a,b), as these terms dominate $\rho$ and $p$. It shows that
dominance of quantum gravity effects may lead to accelerated expansion even
when $\rho >0 ,   \rho \pm p > 0 $ and $\rho + 3 p> 0$ and it is, in absence
of these corrections, decelerated expansion is obtained when these conditions
are obeyed. In the absence of these corrections, acceleration is obtained when
at SEC breaks. It is justified by
early universe model, having dominance of quantum gravity effect and accelerated
expansion. Moreover, in radiation and matter dominated models , $H \sim
t^{-1}$ , so curvature decreases with time giving quantum gravity ineffective
and decelerated expansion for $\rho + 3 p > 0$ as ${\ddot a} < 0$ from
eqs.(10b) and (18a,b).

\bigskip

\centerline {\bf References}
\bigskip

\noindent [1] S. Perlmutter $et$ $al.$, Astrophys. J., {\bf 517}, 565 (1999); A. G. Riess $et$ $al.$, Astron.J., {\bf 116}, 1009 (1998).

\bigskip

\noindent [2] A. D. Miller $et$ $al.$, Astron. J. Lett., {\bf 524}, L1-L4 (1999); P. de Bernardis $et$ $al.$, Nature, {\bf 404}, 955 (2000); S. Hanany $et$ $al.$, Astrophys. J. Lett., {\bf 545}, L5 -L9 (2000); N. W. Halverson $et$ $al.$, Astrophys. J., {\bf 568}, 38 (2002); B. S. Mason $et$ $al.$, Astrophys. J., {\bf 591}, 540 (2003); A. Benoit $et$ $al.$, Astron. Astrophys., {\bf 399}, L25 - L30 (2003).

\bigskip

\noindent [3] D. N. Spergel $et$ $al.$, astro-ph/0302209; L. Page $et$ $al.$,
astro-ph/0302220;  A. G. Riess $et$ $al$, Astrophys. J. {\bf 607}, (2004) 665 [
 astro-ph/0402512].

\bigskip

\noindent [4] R.R. Caldwell, Phys. Lett. B {\bf 545},23 (2002);
astro-ph/9908168 ; R.R. Caldwell, M. Kamionkowski and N.N. Weinberg,
Phys. Rev. Lett. {\bf 91},071301 (2003) ; V.Faraoni, Phys. Rev.D{\bf
68}, 063508 (2003) ; R.A. Daly $et$ $al.$, astro-ph/0203113; R.A. Daly
and E.J. Guerra, Astron. J. {\bf 124},1831 (2002) ;  R.A. Daly,
astro-ph/0212107; S. Hannestad and E. Mortsell, Phys. Rev.D{\bf 66},
063508 (2002); A. Melchiorri $et$ $al.$, Phys. Rev.D{\bf 68}, 043509
(2003); P. Schuecker $et$ $al.$, astro-ph/0211480; H. Ziaeepour,
astro-ph/0002400 ; astro-ph/0301640; P.H. Frampton and T. Takahashi,
Phys. Lett. B {\bf 557}, 135 (2003) ;  P.H. Frampton, hep-th/0302007;
S.M. Carroll $et$ $al.$, Phys. Rev.D{\bf 68} 023509 (2003);  J.M.Cline
$et$ $al.$, hep-ph/0311312; U. Alam , $et$ $al.$, astro-ph/0311364 ,
astro-ph/0403687; O. Bertolami, $et$ $al.$, astro-ph/0402387;
P. Singh, M. Sami $\&$ N. Dadhich, Phys. Rev.D{\bf 68},  023522
(2003); M. Sami $\&$ A. Toporesky, gr-qc/0312009; S. Noriji and
S. D. Odintsov ,Phys.Lett.B,{\bf 562}, 147 (2003)[hep-th/0303117]; Phys.Lett.B,{\bf 571},1(2003)[hep-th/0306212].
\bigskip

\noindent [5] E.J.Copeland, M.Sami and S. Tsujikawa, hep-th/0603057.

\bigskip

\noindent [6] J. D. Barrow, Class. Quant. Grav., {\bf 21}, L79 (2004) [ gr-qc/0403084]; gr-qc/0409062.

\bigskip

\noindent [7] E. Elizalde, S. Noriji and S. D. Odintsov ,Phys. Rev.D{\bf 70}
(2004) 043539 [hep-th/0405034].

\bigskip

\noindent [8] K. Lake , gr-qc/0407107.

\bigskip

\noindent [9] S. Nojiri and S. D. Odintsov , Phys. Lett. {\bf B 595},1 (2004) 
[hep-th/0405078] ; Phys. Rev.D{\bf 70} (2004) 103522 [hep-th/0408170].

\bigskip

\noindent [10] S.K.Srivastava, hep-th/0411221.

\bigskip

\noindent [11] S. Nojiri, S. D. Odintsov and S. Tsujikawa ,
hep-th/0501025.

\bigskip
\noindent [12]I. Brevik,S. Nojiri and S. D. Odintsov , Phys. Rev.D{\bf
70},  043520 (2004)[hep-th/0401073].

\bigskip
\noindent [13]S.K.Srivastava,Phys. Lett. {\bf B 619},1
(2004)[astro-ph/0407048]

\bigskip
\noindent [14] S. Nojiri and S. D. Odintsov , Phys. Rev.D{\bf
68},  123512 (2003)[hep-th/0307288]; hep-th/0601213 .

\bigskip
\noindent [15] S. Nojiri and S. D. Odintsov , Phys. Rev.D{\bf
72},  023003 (2005).

\bigskip
\noindent [16] R.B.Mann, L.Tarasov, D.G.C.Mckeon and T.Stelle , Nucl.Phys. 

\noindent {\bf B 311 },  (1988/89) 630.

\end{document}